# Structural Phases in Non-Additive Soft-Disk Mixtures:
# Glasses, Substitutional Order, and Random Tilings

Asaph Widmer-Cooper

and Peter Harrowell

*School of Chemistry, University of Sydney,*

*Sydney, New South Wales, 2006, Australia*

**Abstract**

Relaxation of the additivity condition on the interaction length between unlike species in a binary mixture of soft disks opens up a rich variety of structures in both crystal and amorphous states with an associated diverse range of relaxation dynamics. We report on MD simulation studies of binary soft disks with negative deviations from additivity that include evidence of accumulation of crystal-like structures in metastable liquids prior to crystallization and the occurrence of a liquid to random-tiling transition.

## 1. Introduction

The filling of space by non-overlapping particles represents the most fundamental of organising principles in condensed matter. Examples of this principle in action are easily found. The requirement that a mixture of hard spheres of two different sizes be densely packed, for example, is sufficient to generate many of the common structures found in atomic crystals [1,2]. The current theory of liquid structure is largely built on the observation [3] that this structure is largely determined by the short ranged steric repulsions. A final example, binary mixtures of purely repulsive spheres [4] and discs [5] have been shown to capture all of the essential phenomenology of the transition from liquid to glass on supercooling. Given this broad utility of mixtures of repulsive particles, it is interesting to consider an extension of the model to allow for chemical ordering (i.e. species-specific association or avoidance) while retaining the geometrical transparency of steric interactions. Non-additivity has the effect of either repelling (for positive deviations) or attracting (for negative deviations) particles of different species. In this paper we shall examine the collective consequences of a deviation from additivity of the interspecies interaction length, i.e. $\sigma_{12} \neq (\sigma_{11} + \sigma_{22})/2$, where $\sigma_{ij}$ is the interaction length between species $i$ and $j$ (described in more detail in Sec. 2).

We shall start with an additive soft disc mixture with $\sigma_{22}/\sigma_{11} = 1.4$, which has been extensively studied [5-7] in the context of the glass transition. With the extension to non-additivity, we shall show that this model can stabilize several substitutionally ordered crystals, structurally different glasses, and a dodecagonal random tiling, without the need for attractive or angle-dependent terms.

This paper is structured as follows. The model and simulation details are described in Sec. 2. In Sections 3 and 4 we consider the effect of a small negative deviation from additivity on the structure and properties of ordered and amorphous phases, respectively. Increasing the negative deviation from additivity produces a square crystal (Sec. 5) and a random square-triangle tiling (Sec. 6). In Sec. 7 we compare and contrast these systems.

## 2. Model and Simulation Details

The binary soft disc model consists of particles interacting via purely repulsive potentials of the form $u_{ab}(r) = \varepsilon \left[ \dfrac{\sigma_{ab}}{r} \right]^{12}$ where $a, b = 1$ or 2. The independent parameters for the model are $\sigma_{22}$, $\sigma_{12}$ and $x_1$, the fraction of particles of type 1. (We have defined the unit length to be $\sigma_{11}$.) We shall fix $\sigma_{22} = 1.4$. In the additive model, this choice would automatically fix $\sigma_{12} = (\sigma_{11} + \sigma_{22})/2 = 1.2$ but for the non-additive system $\sigma_{12}$ can be chosen independently of $\sigma_{11}$ and $\sigma_{22}$. We shall investigate the effect of varying $\sigma_{12}$ from 1.0–1.3. The systems used in this paper consist of $N = 1000$-1500 particles enclosed in a square box with periodic boundary conditions. All units quoted will be reduced so that $\sigma_{11} = \varepsilon = m = 1.0$, where $m$ is the mass of both types of particle. Specifically, the reduced unit of time is $\tau = \sigma_{11}(m/\varepsilon)^{1/2}$ and $T = Tk_B/\varepsilon$. The composition will be specified using $x_1 = N_1/N$, the mole fraction of species 1, the smaller of the particles.

The simulations were carried out at constant pressure ($P = 13.5$) using a generalised leapfrog implementation of the Nosé-Poincaré-Andersen Hamiltonian [8]. Typically, a random configuration was equilibrated at $T = 5$ and then cooled in steps with the equilibration time at each temperature exceeding $100\tau_\alpha$ until either the system crystallized or the time required for this exceeded a maximum equilibration time of $10^4\tau$ and the system became glassy.

## 3. A Small Negative Deviation from Additivity Produces a Chemically Ordered Crystal.

As we are interested in stable mixtures, our focus in this paper will be on negative deviations from additivity. The instability associated with a *positive* deviation (i.e. $\sigma_{12} > (\sigma_{11} + \sigma_{22})/2$) is easily demonstrated. Upon cooling below $T = 2$, the equimolar mixture with $\sigma_{12} = 1.3$ segregates into small and large particle-rich liquid fractions (see Fig. 1b). On further cooling, the separated liquids freeze at different temperatures; first, the large particle component at $T_{f,2} \approx 1.3$ (Fig. 1c) and then the small particle component at $T_{f,1} \approx 0.8$ (Fig. 1d). This is visible as a rapid increase in the relaxation times for the large and small particles, respectively. These freezing temperatures are substantially reduced relative to the freezing temperatures obtained for the respective one-component systems ($T_{f,2} = 1.7$ and $T_{f,1} = 0.95$) [5].

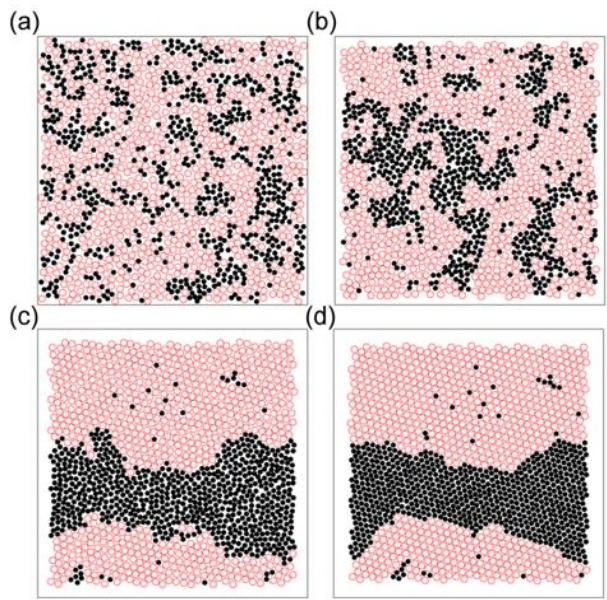

Figure 1. Particle configurations for the equimolar mixture with $\sigma_{12} = 1.3$ at (a) $T = 3$, (b) $T = 2$, (c) $T = 1.2$, and (d) $T = 0.4$, showing the four different phase regions. Species 2 (the larger particles) are indicated by open circles.

Making $\sigma_{12}$ smaller than $(\sigma_{11} + \sigma_{22})/2$ stabilizes the mixture by introducing an effective attraction between unlike particles species. At $\sigma_{12} = 1.1$ (an 8% reduction from the additive value $\sigma_{12} = 1.2$) we find a dense packed substitutionally order crystal at the equimolar composition. This AB crystal, shown in Fig. 2, was first described by Likos and Hanley [9] who labelled it H2. Kennedy [10] has subsequently confirmed that the H2 structure represents the compact packing of (additive) hard disks with a radius ratio of 0.637556. Hentschel et al [7] have proposed the H2 structure as the ground state of the additive mixture when constrained to be compositionally homogeneous.

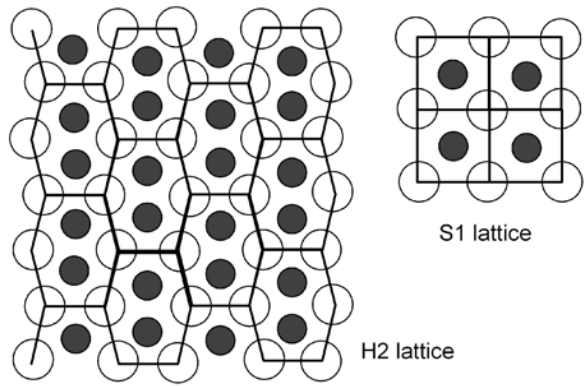

Figure 2. An ideal H2 crystal lattice and the related S1 lattice. Species 2 (the larger particles) are indicated by open circles.

The H2 crystal structure can be thought of as the result of a periodic distortion of the more symmetric square lattice (S1) structure (right). The transition between the liquid mixture and the H2 crystal is first order, as evidenced by the density change and hysteresis evident in Fig. 3. The depth of the supercooling achievable for this mixture is substantial. We shall return to the question of the kinetic stability of the metastable liquid below.

A sequence of configurations taken at $T = 0.6$ during the freezing transition are shown in Fig. 4. Nucleation and growth are not observed until the metastable liquid has been equilibrating for more than $110\tau_{\alpha}$, where $\tau_{\alpha}$ is the time required at $T = 0.6$ for the self intermediate structure factor of the large particles $F_{s,2}(k_2,t)$ to decay to 0.1 (here $k_2 = 5.8$ is the Bragg wavevector and corresponds to the position of the first peak in the large particle structure factor). The density of this defective crystal is indicated by the black diamond in Fig. 3 and contains both parallel and herringbone arrangements of the H2 unit (see Fig. 8 for examples). The structure of the growing crystal appears to be quite labile; some of the herringbone packing present during crystallisation has changed to parallel packing by the time the main crystallisation event is complete (Fig. 4). We note that because the H2 cells are able to pack in mixed parallel and herringbone alignments, it is possible to have perfect H2 local order without complete long-range translational order.

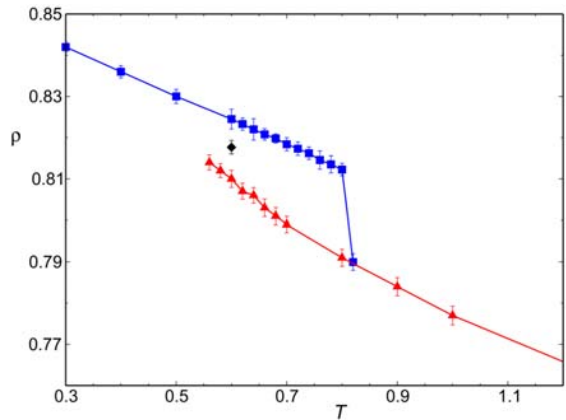

Figure 3. Isobars for the liquid and crystal through the transition temperature for the equimolar mixture with $\sigma_{12} = 1.1$. Blue squares and red triangles indicate data points of the heating and cooling traverses, respectively. The black diamond indicates the final density after a freezing transition at $T = 0.6$. Error bars represent one standard deviation. Note the large region of metastability extending from $T = 0.6$ to $0.8$. We estimate the equilibrium transition temperature to lie between $T = 0.7$ and $0.8$.

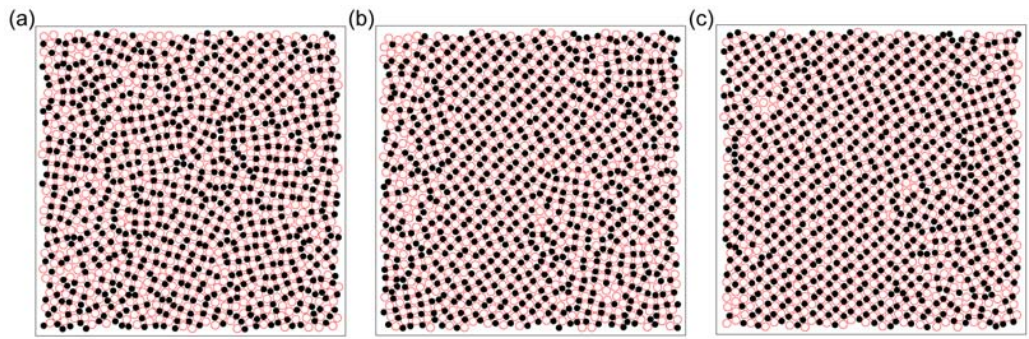

Figure 4. Configurations for the equimolar mixture with $\sigma_{12} = 1.1$ at $T = 0.6$ (a) before, (b) during, and (c) after freezing. The times are $75\tau_{\alpha}$, $225\tau_{\alpha}$, and $450\tau_{\alpha}$ since the start of equilibration.

Approaching the freezing temperature from above, we observe that the liquid accumulates considerable local and medium range order, similar to that found in the crystal, prior to the phase transition. Local structure about the small and large particles are indentified as S$n_s n_l$ and L$n_s n_l$, respectively, where $n_s$ and $n_l$ are the number of small and large neighbours, respectively. Fig. 5 shows the average fraction of small and large particles with various local packing environments as a function of temperature. (We have used the position of the first minimum in the appropriate partial radial distribution function $g_{ab}(r)$ to identify nearest neighbors.) As the freezing point is approached, we observe the steady increase in the fraction of L43 and S14 environments (the two environments found in the crystal). By the time freezing occurs at $T = 0.6$, over half the particles are in these environments.

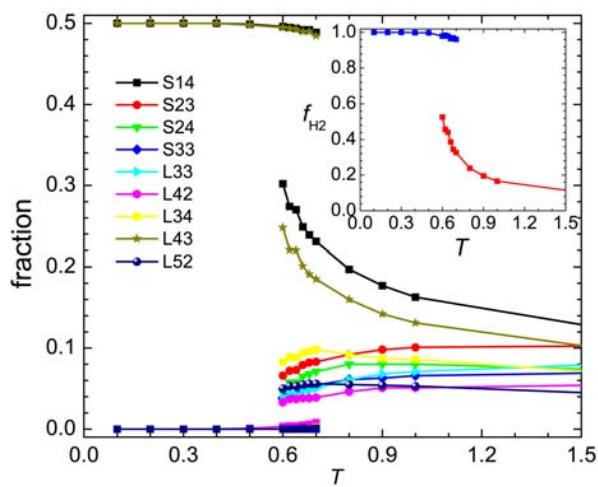

Figure 5. The distribution of local packing environments during the heating and cooling traverses for the equimolar mixture with $\sigma_{12} = 1.1$. (Inset) The fraction of particles in H2 unit cells $f_{H2}$.

It is possible to accumulate a significant concentration of S14 local arrangements without having anything resembling the crystal phase since the H2 cell requires a pair of overlapping S14 environments. In this case, however, we find that the H2 cell is substantially present in the liquid. Fig. 5 shows that, at the temperature at which the freezing transition occurred, over half of the particles in the (metastable) liquid belonged to an H2 cell. In fact, the fraction of particles in an H2 cell, $f_{H2}$, is well fit over the temperature range $0.6 \leq T \leq 5$ by the power law $f_{H2} = A \, (T - T_o)^{-\gamma}$ with $T_o = 0.460$, $\gamma = 0.872$ and $A = 0.0947$. If this tend continued, the fraction of crystalline order in the liquid would become 1 at $T = 0.527$. Despite this local organization, the liquid exhibits neither long-range translational nor orientational organization of the local H2 cells (see Fig. 6). Below $T = 0.7$, the liquid structure is characterised by clusters of H2 cells in both parallel and herringbone packing that are 3–4 H2 units across but randomly oriented with respect to one another.

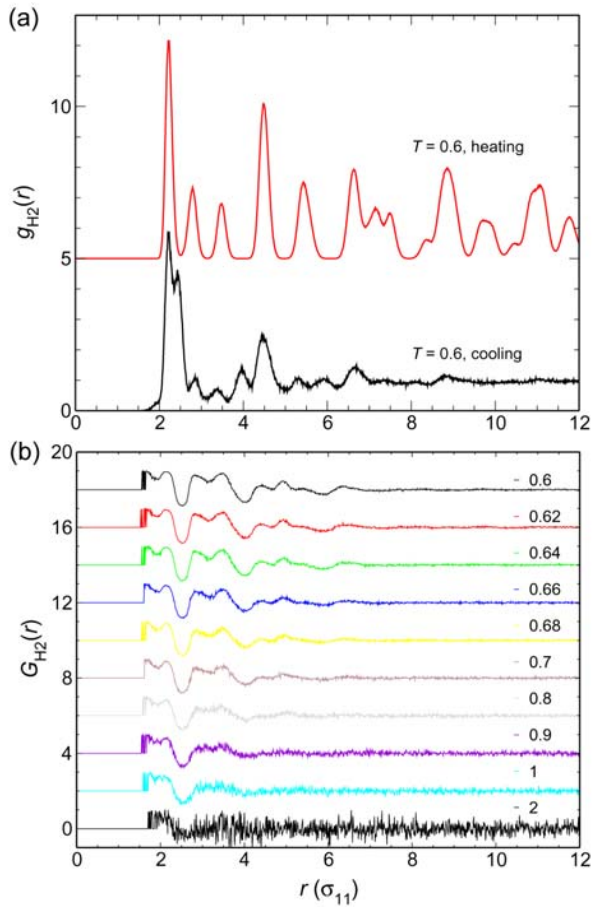

Figure 6. (a) The pair distribution function for H2 cells $g_{H2}(r)$ at $T = 0.6$ during the cooling and heating traverses. (b) The temperature dependence of the two-fold orientational correlation function for H2 cells $G_{H2}(r)$ during the cooling traverse. For clarity, functions have been offset vertically from each other. See the Appendix for the definitions of $g_{H2}(r)$ and $G_{H2}(r)$.

The equimolar mixture with $\sigma_{12} = 1.1$ provides an instructive paradox: a liquid that accumulates considerable local structure, crystal-like no less, but that still exhibits sufficient stability with respect to crystallization to support supercooling to ~75% of the melting point. The single component soft disk liquid also develops crystal-like local structure before freezing [11], but shows no such stability.

The rate-limiting step for crystal nucleation and growth of the non-additive mixture appears to be alignment of H2 cells in the form of a coherent cluster greater than some critical size. Evidence for this can be found in Fig. 7, where we compare the time dependence of the fraction of particles in H2 cells in a crystallizing run with the evolution of the distribution of orientations of these H2 cells. Continuous growth of the H2 fraction only occurs after the appearance of the twin peaks in the orientation distribution, separated by the $80^o$ that is indicative of the herringbone structure (Fig. 8a). (There are signs at the longest times of one of these two peaks diminishing indicating a slow conversion to the parallel structure shown in Fig. 8b).

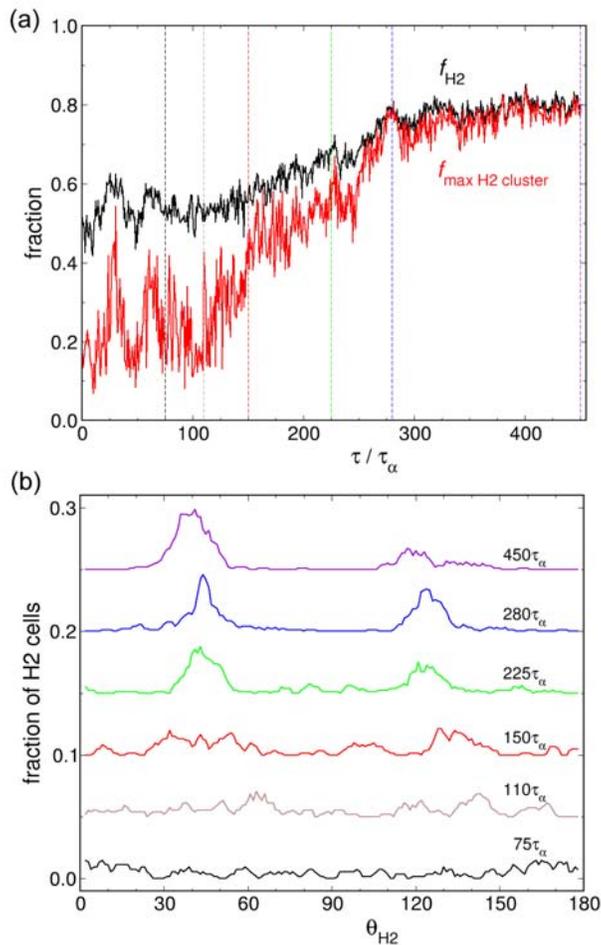

Figure 7. Structural changes prior to and during crystallization for the equimolar mixture with $\sigma_{12} = 1.1$ at $T = 0.6$: (a) The fraction of particles in H2 order and in the largest cluster of H2 cells, and (b) the angular distribution of H2 cells at times corresponding to the dashed lines in (a).

Crystallization in the run described in Fig. 7 does not start until the system has been equilibrating for >110$\tau_\alpha$, after which the density increases for ~185$\tau_\alpha$, and the crystal continues to anneal over even longer times. In comparison, the average particle diffuses 1.7-2.6 diameters in 75$\tau_\alpha$. The high degree of crystalline order in the liquid prior to freezing certainly reduces the magnitude of the extensive changes on crystallizing: the volume decreases by only ~1% upon freezing and the change in potential energy (not shown) is barely noticeable relative to the magnitude of its fluctuations. Non-crystalline packings of the H2 cells, such as that sketched in Fig. 8c, represent stable obstacles to the global organization of the local order.

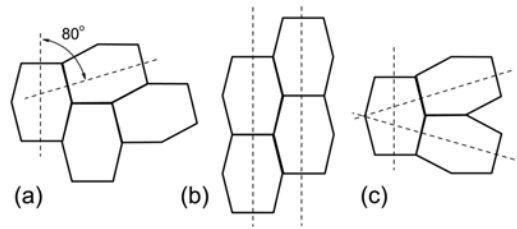

Figure 8. Packings of H2 cells: (a) herringbone, (b) parallel, and (c) non-crystalline.

## 4. A chemically ordered glass

Slow liquid kinetics, resulting from the accumulation of local order, could lead to glass formation. Exploring this possibility in the $\sigma_{12} = 1.1$ mixture simply requires that we vary the composition away from that of the H2 crystal. Here we shall study the behaviour of a mixture with a species 1 mole fraction $x_1 = 0.317$. (This specific composition was chosen because it is related to the entropy maximum in the $\sigma_{12} = 1.0$ mixture we describe in the following section.) In contrast to the equimolar mixture, we see no evidence of crystallization. In Fig. 9, we plot the fractions of the various local environments. There is little change in structure with temperature save for what appears to be a closely correlated increase in S05 and L34. (The reader is reminded that the H2 crystal is comprised of S14 and L43 environments.) So, with the shift in composition, we find a smaller increase in local order on cooling and what does form is a local order quite distinct from that of the crystal. We have not made a detailed study, but these observations clearly show that the liquid structure undergoes significant transformation as the composition is changed. As described below, the decrease in $x_1$ appears to result in a 'dissociation' of the pairs of small particles characteristic of the H2 cell. The nature of this dissociation represents an interesting problem for the future.

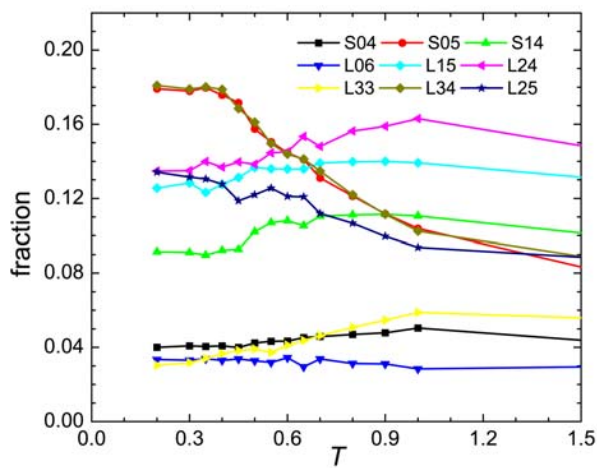

Figure 9. The distribution of local packing environments as a function of temperature for $\sigma_{12} = 1.1$, $x_1 = 0.317$. We have identified a particular neighbourhood with the following notation: A small particle with $n_S$ small neighbours and $n_L$ large neighbours is designated as $Sn_Sn_L$ and the analogous large particle is indicated as $Ln_Sn_L$. Only the most common environments are shown.

In the $x_1 = 0.317$ mixture at $T = 0.2$, five environments make up 75% of the local structure and four environments contribute a further 20%. Furthermore, these environments are so distributed in space so as to produce a quite homogeneous appearance to the phase (see Fig. 10). In contrast, the glassy state of the equimolar mixture of additive disks [5,6] is quite heterogeneous with transient domains of single component crystals. On a local level, the additive mixture is also more diverse, with over 20 different local environments, none of which contribute more than 10% of the total. As the structures of atomic (including ionic) crystals rarely run to more than 4 distinct local environments [12], it is tempting to speculate that a condensed phase with more than some threshold number (say 4-6) of stable local environments must be amorphous. If that is so, then the glassy state of our $x_1 = 0.317$ mixture might well lie close to this crossover.

The significant increase in the S05 and L34 environments appears, from inspection of configurations (see Fig. 10), to be a consequence of the packing of the pentagonal S05 units together with triangular units of three large particles and, occasionally, square S04 units and pentagonal S14 units. These local structural units and some of their more common packings are illustrated in Fig. 11. Observe that the large particles at the centre of (b) and (d) both have L34 local environments. Occasionally an H2 cell, consisting of two overlapping S14 environments (a), replaces the square cell in structure (d).

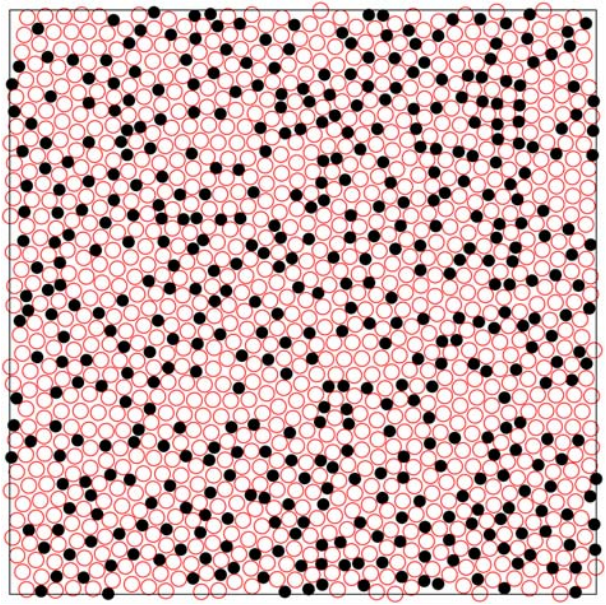

Figure 10. Configuration at $T = 0.2$ showing the scarcity of local crystalline order for $\sigma_{12} = 1.1$, $x_1 = 0.317$.

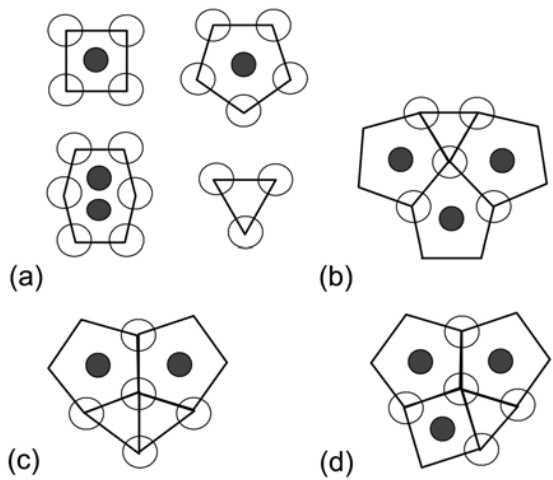

Figure 11. (a) Small particle environments in the low-temperature liquid, and (b)-(d) some common ways in which these pack together to form vertices about large particles. The filled and open circles indicate small and large particles, respectively. Note that the large particles at the centre of (b) and (d) both have L34 local environments.

The preference for the S05 environment over the S14 environment is a bit of a puzzle. When all interparticle distances are set equal to $\sigma_{ab}$, the S14 environment has an angle sum about the small particle of 363°. In contrast, the S05 environment has an angle sum of 395° and, therefore, represents the poorer packing. A key feature of the global structure of the low $T$ phase, as depicted in Fig. 12, is the isolation of the small particles made possible by the decrease in $x_1$ from 0.5 to 0.371. Taking full advantage of the smaller length afforded by the non-additive value of $\sigma_{12}$ to increase density, the particles appear to be organised to avoid contact between small particles. The relative predominance of the S05 environments is the consequence. The large particles are completely employed in maintaining this separation of the small particles, hence the absence of L06 environments in a system whose groundstate is expected to consist of a mixture of the H2 crystal and a crystal of large particles only.

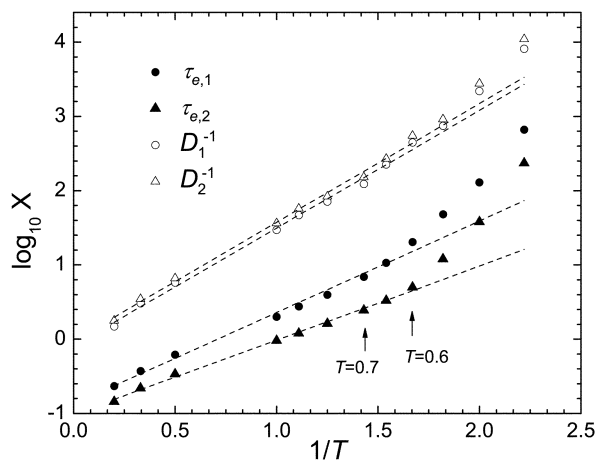

Figure 12. Arrhenius plot of the structural relaxation times $\tau_{e,1}$ and $\tau_{e,2}$ and the inverse diffusion constants $D_1^{-1}$ and $D_2^{-1}$ for $\sigma_{12} = 1.1$, $x_1 = 0.317$. The dashed lines are linear regressions through the data for $T \geq 0.65$.

The $x_1 = 0.371$ non-additive mixture starts to show glassy dynamics below $T = 0.65$: the relaxation times deviate from Arrhenius behavior (Fig. 12) and the non-Gaussian factor $A(t)$ develops a significant peak that moves to longer times [13]. Below $T = 0.45$, the self intermediate scattering functions $F_s(k,t)$ are no longer able to decay to zero within the finite time scale of the simulations. Interestingly, the structural relaxation time $\tau_{e,2}$ from $F_s(k,t)$ calculated for the large particles only is substantially smaller than $\tau_{e,1}$ ($\tau_{e,i}$ is the time required for $F_s(k,t)$ for species $i$ to decay to $1/e$). This is unusual as it means that the large particles are able to relax their local structure faster than the small particles. It is, however, consistent with our proposal above that the stability of the amorphous phase derives from the constraint of non-contact of the small particles, this increasing their effective radius. In contrast to the structural relaxation times, we find little difference between the long-time diffusion constants of the two species.

We note that this model has already been used in several recent publications [14] because of its strong stability to coarsening, which allows for the study of long trajectories and the use of rare event sampling methods.

## 5. A Large Negative Deviation from Additivity Generates Square-Triangle Tilings

If we continue to decrease the inter-species distance $\sigma_{12}$ to 1.0 (a reduction of 20% of the additive length), we find that crystallization of the equimolar mixture becomes facile with little hysteresis observed on heating. The crystal structure consists of two interpenetrating square lattices, each consisting of one species. The resulting square lattice structure has been labelled S1 by Likos and Henley [9] and is depicted in Fig. 13.

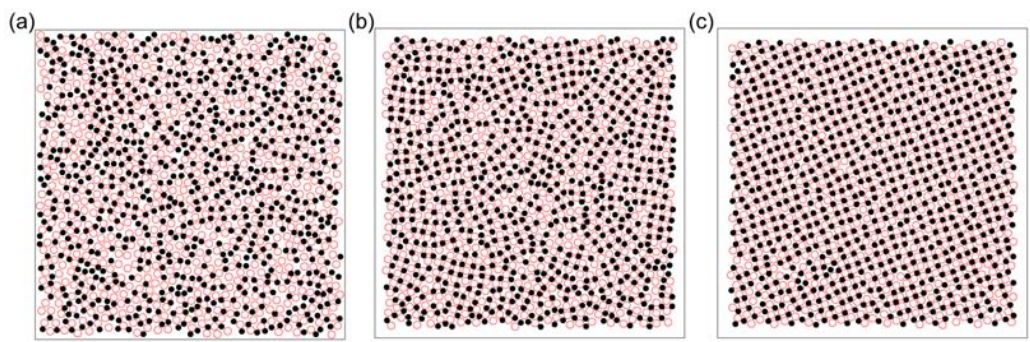

Figure 13. Representative particle configurations for the equimolar mixture with $\sigma_{12}$ = 1.0 at (a) $T$ = 2, (b) $T$ = 1.04, and (c) $T$ = 1.02 during the cooling traverse.

Crystallization of the equimolar mixture is associated with a small but abrupt increase in the density at $T$ = 1.02, as shown in Fig. 14. When an ideal S1 crystal is heated there is a clear step decrease in density at $T$ = 1.08, the difference marking the extent of the hysteresis.

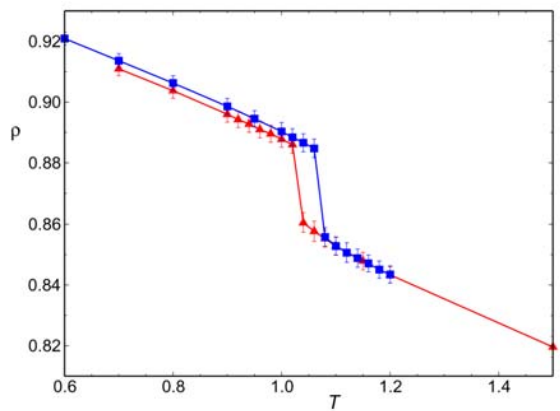

Figure 14. Density vs *T* for the equimolar mixture with $\sigma_{12} = 1.0$. Squares (blue) and triangles (red) indicate data points for the heating and cooling traverses, respectively. Error bars represent one standard deviation.

The square lattice structure of the S1 crystal experiences large amplitude fluctuations. As shown in Fig. 15, the dominant local environments of the crystal over a broad range of temperatures includes both the expected S04 but also the distorted S14. As the melting point is approached, these two environments appear with close to equal frequency. The large particle environments show no such fluctuations with the expected L44 the only dominant environment right up to melting. In the liquid at *T* = 1.08, the S04 and S14 packings account for about 30% of all particles and the L44 and L34 packings account for another 26%. This suggests that about half of all particles are in local crystal-like environments just prior to freezing, which is similar to what we observed for the equimolar mixture with $\sigma_{12} = 1.1$. Analysis of the cooling traverse gives similar results. That the $\sigma_{12} = 1.0$ mixture crystallizes so much more quickly than the $\sigma_{12} = 1.1$ mixture is a consequence of the higher symmetry of the local square structure in the former. The square structure avoids the non-crystalline packings that impede crystallization in the $\sigma_{12} = 1.1$ mixture.

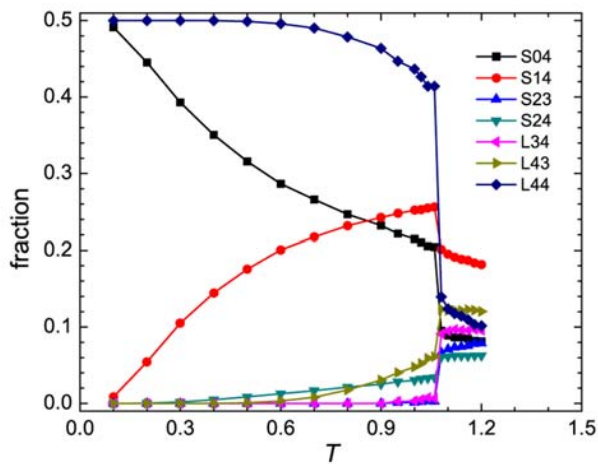

Figure 15. The distribution of local environments as a function of temperature for the heating traverse of the equimolar mixture with $\sigma_{12} = 1.0$.

## 6. Random tilings and the glass state.

When $\sigma_{12} = 1.0$, order at equimolar composition is associated with the square S1 environment while the local order for the single component liquid is characterised by triangular packing. As we vary the composition away from the equimolar value, we see configurations comprised, almost completely, of square S04 units and equilateral triangular units of 3 large particles.

As sketched in Fig. 16, the square and the triangle can be assembled in a number of basic combinations capable of tiling space without defects with the ratio of squares to triangles being determined directly by the composition. The random tiling of the plane by squares and triangles has been the subject of extensive studies [15,16]. The maximum configurational entropy occurs at a ratio of squares to triangles of $\sqrt{3}/4$ [15], which corresponds to the composition $x_1 \sim 0.317$ that we have already made use of in this paper. While a random tiling does not exhibit translational order, it does have a 12-fold orientational order and so is not an isotropic phase. Here we show that thermal fluctuations, in the form of point defects (i.e. local structures other than those depicted in Fig. 16), are sufficient to destroy this long-range orientational order.

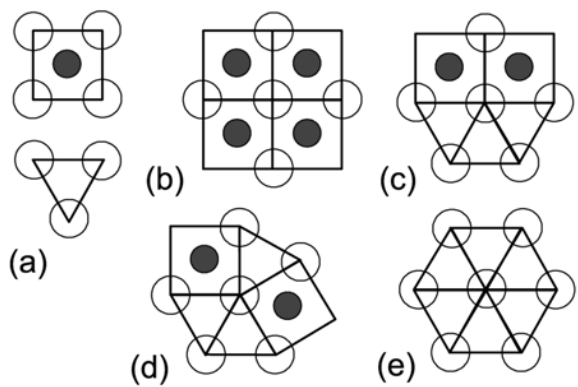

Figure 16. (a) Small particles (filled circles) and large particles (open circles) pack together to create square and triangular tiles (thick lines). These units can further pack together to form the local configurations around large particle vertices that are shown in (b)–(e). The central large particle and its nearest neighbours have been indicated to emphasise the local coordination environment.

On cooling, the $x_1 = 0.371$ mixture (with $\sigma_{12} = 1.0$) undergoes a continuous transition from a liquid to a defected random tiling and the density increases smoothly. Selected particle configurations are plotted in Fig. 17. At $T = 1$ the liquid appears homogeneous but by $T = 0.7$ local square and equilateral packings are apparent and small regions of S1 crystal order can be observed. By $T = 0.4$ the structure can be well described as a random tiling of squares and equilateral triangles with the occasional defect. The low-temperature relaxation appears to consist mainly of the motion of these defects, which involves the displacement of both small and large particles and explains why the structural relaxation times of the two species are almost identical (see below).

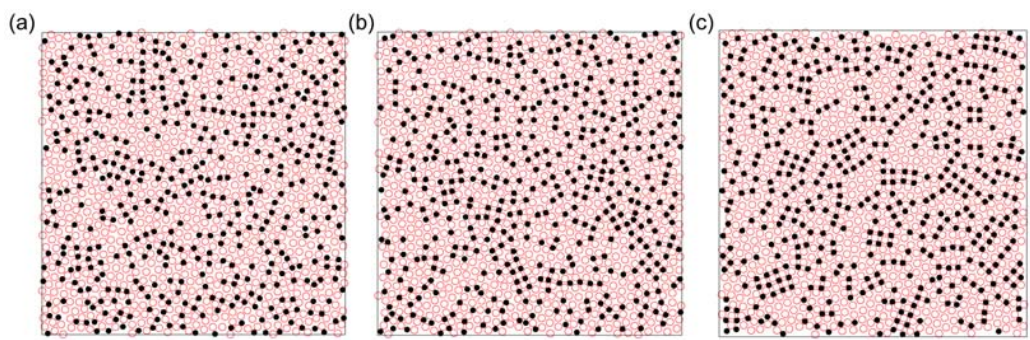

Figure 17. Representative particle configurations for the $x_1 = 0.317$ mixture with $\sigma_{12} = 1.0$ at (a) $T = 1$, (b) $T = 0.7$, and (c) $T = 0.4$ during the cooling traverse.

Fig. 18 shows the distribution of local environments as the liquid is cooled. The transition in the structure is broadened out over a range of temperatures. The main change is a large increase in the fraction of L25 packings, which correspond to particles at the centre of vertices of types (c) and (d) shown in Fig. 16. This is the dominant large-particle environment at low temperature. The L06 packing, corresponding to vertex (e), contributes 6%, and the L44 packing, corresponding to vertex (b), accounts for less than 3.5% of the total. Thus the local large-particle order is substantially different from the equimolar mixture. In total, at least 68% of the large particles are in square-triangle tiling environments at low temperature. The substantial fraction of S14 order at low temperature must be due to a vibrational distortion of two S04 units packed together. Counting both S04 and S14 environments, this suggests that about 88% of the small particles are in square-triangle tiling environments at low temperature. This leaves us with a high fraction of particles in defects sites, roughly 37%.

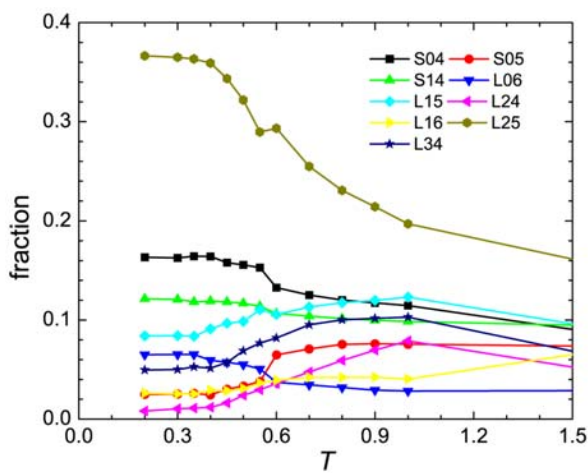

Figure 18. The distribution of local packing environments for the $x_1 = 0.317$ mixture with $\sigma_{12} = 1.0$ as a function of temperature.

The various relaxation times (see Fig. 19) show no abrupt change on cooling but rather increase in a manner similar to the standard scenario for a glass transition of a supercooled liquid. This observation is consistent with the continuous nature of the structural transition already noted. We are left, then, with the interesting prospect of what appears to be a bona fide glass transition but to a glassy phase characterised by extensive local order (random tiling) and well-defined structural defects. Since relaxation is dominated by the movement, creation and annihilation of these defects, it should be possible to provide an explicit account of how the temperature dependence of the dynamics is related to the microscopic reorganizations associated with these defects [17].

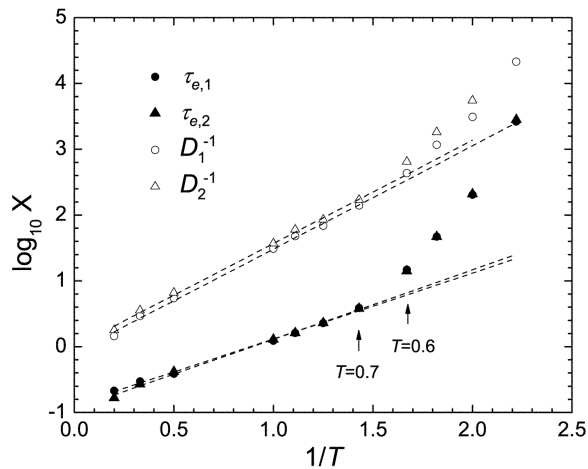

Figure 19. Arrhenius plot of the structural relaxation times $\tau_{e,1}$ and $\tau_{e,2}$ and the inverse diffusion constants $D_1^{-1}$ and $D_2^{-1}$ for the $x_1 = 0.317$ mixture with $\sigma_{12} = 1.0$. The dashed lines are linear regressions through the data for $T \geq 0.7$. Note the divergence from Arrhenius behaviour at low temperature.

## 7. Discussion

In this paper we have described how a modest change of the cross-species interaction length away from additive can significantly change the degree of chemical ordering and the geometric consequences of this ordering for a binary mixture of soft disks. Fig. 20 provides a summary of the phase behaviour described in this paper.

We have demonstrated that the additive binary mixture with $\sigma_{22}/\sigma_{11} = 1.4$, studied extensively as being a good glass former, sits at something of a crossover point between demixing (when $\sigma_{12}$ is increased to 1.3) and substitutionally ordered crystallization (when $\sigma_{12}$ is decreased to 1.1). Negative deviations from non-additivity produce both an increase in the association between unlike species and changes in the locally and globally preferred structures. In this study we have explored the consequences of this coupling for both crystallization and glass formation.

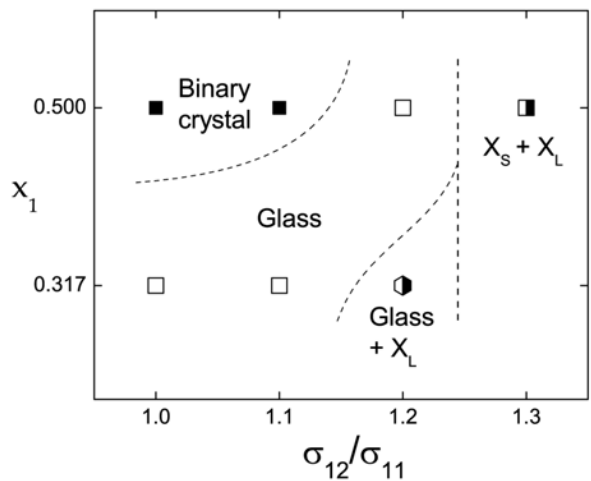

Figure 20. Overview of phase behavior as a function of composition and non-additivity for the binary soft-disk model. $X_S$ and $X_L$ indicate hexagonal crystals of pure small or large particles, respectively. The dashed lines are guides only.

Let us consider the scenarios for crystallization and glass formation associated with the three choices of non-additivity reported here. We shall begin with the case of $\sigma_{12} = 1.3$. Previously [4], we have argued that the equilibrium state for the additive mixture at low temperatures consists of the phase separated crystals. By increasing $\sigma_{12}$ from 1.2 to 1.3, we have decoupled the demixing process from the crystallization so that the separation of species can occur while both phases are still liquid and, hence, quickly. The subsequent crystallization of each of these resulting pure phases on further cooling occurs with little to no nucleation barrier. If we were to reduce the magnitude of the positive deviation from additivity (i.e. select $1.2 < \sigma_{12} < 1.3$) the temperature at which the species segregation occurs must eventually drop below the freezing point of the pure liquid of large particles and the segregation process becomes coupled, thermodynamically and, more importantly, kinetically, to solidification. This coupling results in a larger activation barrier for crystallization. This, then, is the scenario by which the glass transition of the additive mixture is achieved – crystallization and segregation are hitched together so that one cannot occur without the other.

Shifting now to the negative deviation from additivity and $\sigma_{12} = 1.1$, segregation is replaced by chemical association of unlike species. At $x_1 = 0.5$, this chemical ordering results in the H2 cell whose stability is such that the structure accounts for half the particles in the equilibrium liquid at the freezing point. The rate determining processes of crystallization is not the generation of crystal-like structure but rather the alignment of the structure already present. Nucleation in this system is not measured in terms of the local order but rather the intermediate order sensitive to the alignment of the individual H2 cells. This is a scenario analogous to the fluctuation-induced first order transitions considered by Brazovskii [18].

The H2 geometry is restricted to compositions close to equimolar. The pair of small particles that comprise the core of the cell are there as a compromise between packing particles efficiently and reducing contact between like particles. On decreasing $x_1$, these paired particles and the H2 cell that housed them vanish, to be replaced by single small particles encased in a pentagon of large ones. This local structure does not form part of any of the compact crystal structures of mixtures of disks. This second glass-forming scenario perhaps most closely resembles the popular conception of a supercooled liquid that is stabilised by the frustration arising from locally favored structures that are inconsistent with any periodic crystal structure [19].

Finally, we come to the $\sigma_{12} = 1.0$ mixture and the last scenario for crystallization and glass formation. Following the trend set by the $\sigma_{12} = 1.1$ liquid, the equimolar mixture crystallizes into a substitutionally ordered solid. What is new is the rapidity of the crystallization with little hysteresis observed. The accumulation of local order in the liquid does not result in the associated slowing down observed in the previous model. We can tentatively attribute the ease of crystallization to two features of the local order. First, the higher symmetry of the square S1 cell eliminates the possibility of non-crystalline arrangements. Second, the square S1 cell is soft and subject to substantial fluctuations, as evidenced by the substantial population of S14 environments. Such fluctuations should assist in structural relaxation.

The random square-triangle tiling state, observed in the $\sigma_{12} = 1.0$ mixture for $x_1 < 0.5$, represents a remarkable merging of crystallization and glass formation. While the perfect random tiling phase is not isotropic and, therefore, not a glass, thermal excitation of defects can render the state isotropic and the associated transition, given its continuous character, indistinguishable from a glass transition. This liquid provides a striking illustration of what can happen when the favoured local order can be organised into a large multiplicity of degenerate groundstates. If we accept that the transition resembles that from liquid to glass then we are left with an example of a glass characterised by a simple local order (the square-triangle combinations of Fig. 16) and well-defined defects via which relaxation proceeds. This glass transition is then a relatively straightforward process to explain – i.e. a disorder to order transition where the excitation energy in the ordered state is low.

Both the crystallization and glass transitions that we have characterized are preceded by an accumulation of local structure in the equilibrium liquid. The observation for the non-additive mixtures that a significant amount of the liquid structure can be associated with specific favoured environments and that these environments can include structure found in the crystal matches similar observations reported recently for a binary mixture of Lennard-Jones particles [20] and a lattice model of liquids with low symmetry favoured local structures [21]. These results challenge the notion that liquids lack well defined order or that crystal-like structures can play no role in liquid behaviour, except as unstable modes leading to crystallization in the supercooled state.

## 8. Conclusion

We have studied the effect of non-additivity on the low-temperature phase behaviour and kinetics of liquids comprised of a binary mixture of soft disks. Non-additivity allows us to introduce chemical ordering (or drive chemical segregation) in a liquid governed by only steric interactions. Making only modest changes to a single length scale in the model, the interspecies interaction length $\sigma_{12}$, we have demonstrated a rich variety of local structures and their associated collective behaviour. Perhaps most striking, we have identified three distinct scenarios for glass formation, i.e. coupling compositional and crystalline fluctuations, stabilizing local structures that are incompatible with crystal structures and, finally, achieving a dense non-crystalline arrangement of favoured local structures through the presence of a random tiling (with defects). The possibility that the glasses associated with these different scenarios represent distinct classes of amorphous materials is an interesting question for future study.

**Appendix: Definition of $g_{H2}(r)$ and $G_{H2}(r)$**

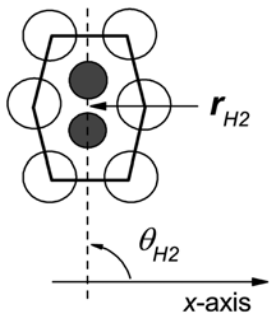

Figure 21. Definitions of $\boldsymbol{r}_{H2}$ and $\theta_{H2}$ used to calculate $g_{H2}(r)$, $G_{H2}(r)$, and the angular distribution of H2 cells in Sec. 3.

Using the definitions of $r_{H2}$ and $\theta_{H2}$ illustrated in Fig. 21, we define a pair distribution function for H2 cells as

$$g_{H2}(r) = \left\langle \frac{1}{N_{H2}\rho_{H2}} \sum_{i=1}^{N_{H2}} \sum_{j \neq i}^{N_{H2}} \delta(r - r_{H2,ij}) \right\rangle$$

where $N_{H2}$ is the total number of H2 cells, $\rho_{H2} = V/N_{H2}$, $r_{H2,ij} = |\boldsymbol{r}_{H2,i} - \boldsymbol{r}_{H2,j}|$ is the separation between the H2 cells $i$ and $j$, and the angular brackets denote an average over different configurations in time. To test for the presence of long-range orientational correlation between H2 domains we define an orientational correlation function for H2 cells as

$$C_{H2}(r) = \left\langle \frac{1}{N_{H2}\rho_{H2}} \sum_{i=1}^{N_{H2}} \sum_{j \neq i}^{N_{H2}} \cos[2(\theta_{H2,i} - \theta_{H2,j})]\delta(r - r_{H2,ij}) \right\rangle$$

where the angular brackets again indicate an average over time origins. This orientational correlation function is weighted by the translational correlations. To see it free of this bias we plot the ratio $G_{H2}(r) = C_{H2}(r)/g_{H2}(r)$. If all the H2 cells at a given separation r lie parallel to each other then $G_{H2}(r) = 1$, while if they all lie perpendicular to each other then $G_{H2}(r) = -1$.

**ACKNOWLEDGEMENTS**

This work was supported by funding from the Australian Research Council. AW acknowledges the support of the Australian Research Council in the form of an Australian Postdoctoral Fellowship.